\documentclass[12pt]{article}

\def\be{\begin{equation}}
\def\ee{\end{equation}}
\def\ba{\begin{eqnarray}}
\def\ea{\end{eqnarray}}
\def\half{{1 \over 2}}
\def\nave{n_{\rm ave}}
\def\kBo{k_{\rm Bo}}
\def\e{\mbox{e}}
\def\k{{\bf k}}

\def\x{{\bf x}}
\def\lmax{\lambda_{\max}}
\def\lphi{\lambda_{\phi}}
\input epsf
\begin{document}
\title{Strong acoustic turbulence and the speed of Bose-Einstein condensation}
\author{S. Khlebnikov \vspace{0.1in} \\
{\it \normalsize Department of Physics, Purdue University, West Lafayette, 
IN 47907, USA}}
\date{January 2002}
\maketitle
\begin{abstract}
The final stage of Bose-Einstein condensation in a large volume
occurs through coarsening---growth 
of individual correlated patches. We present analytical arguments and
numerical evidence that in the momentum space this growth
corresponds to strong turbulence of the particle number, with a turbulent
cascade towards the infrared and the power law $n(k) \propto k^{-(d+1)}$ in 
$d=2,3$ spatial dimensions. As a corollary, we find that the correlation 
length grows linearly in time, with the speed of order of the speed of 
Bogoliubov's quasiparticles (phonons). We use these results to estimate the
speed of BEC in atomic vapors and in galactic dark matter. \\
\hspace*{3.9in} hep-ph/0201163
\end{abstract}
\section{Introduction}
The discovery of Bose-Einstein condensation (BEC) in alkali vapors 
\cite{exp1}--\cite{exp3} had stimulated
interest in a theoretical description of non-ideal Bose gases, a research topic
that dates back to Bogoliubov's work \cite{Bogoliubov}
on the equilibrium properties.
In current experiments, the gas is confined in a small magnetic trap and is strongly
affected by the confining potential. However, theorists have also investigated
the kinetics of BEC in a uniform (on average) gas, contained
in a macroscopic volume \cite{Kraichnan}--\cite{grinst}. 
In addition to the intrinsic interest that this
problem has for basic science, it is also of potential importance for astrophysics
and cosmology in connection with the problem of galactic dark matter 
\cite{it86}--\cite{Riotto&Tkachev}.

The earliest theoretical work we are aware of on this subject
is the paper by Kraichnan \cite{Kraichnan}. Although we will confirm below
several qualitative results of that work 
(such as concentration 
of particles in a progressively smaller momentum range and a turbulent cascade at 
larger momenta), the quantitative details differ. We attribute the difference 
to the imprecise treatment of phonons in ref. \cite{Kraichnan} (noted by Kraichnan
himself). Although the amount of theoretical work on BEC in a macroscopic volume 
has grown rapidly in recent years \cite{KSS}--\cite{grinst}, an adequate understanding
of the phonon dynamics has remained elusive. In the present paper, we
attempt to improve the situation, through a combination of analytical estimates and
numerical results.

We consider an isolated nonrelativistic Bose gas, 
sufficiently dilute for the 
interaction between the particles (atoms in the laboratory, or dark
matter particles in the outer space) to be characterized by 
a single parameter---the scattering length. Such a gas can be described by an
effective Hamiltonian:
\be
H = \frac{\hbar^2}{2m} \nabla \psi^{\dagger} \nabla\psi
+ \half g_4 (\psi^{\dagger} \psi)^2 \; ,
\label{H}
\ee
where $\psi$ and $\psi^{\dagger}$ are canonically conjugate (field) variables,
$m$ is the particle mass, and $g_4$ is the coupling that encodes the value of
the scattering length.\footnote{
In the context of dark matter, using (\ref{H}) implies neglect of gravitational
interactions, which in general is only possible up to some spatial scale. 
We discuss this question further in the conclusion.}
In a macroscopic volume, a negative $g_4$ 
(i.e. an attractive interaction) leads to clumping, the process that had been
investigated numerically in ref. \cite{dew}. In the present paper we consider
the case of a positive $g_4$ (a repulsive interaction). In 
the concluding section, we will also discuss briefly the case when the coupling 
is switched from positive to negative, as in a recent experiment \cite{collapse}.  
We assume that the volume is macroscopic, and the gas is on average uniform.

A fundamental role in the kinetics of a Bose gas is played by the length scale
corresponding to the wavenumber $k_{\rm Bo}$,
\be
\hbar^2 k_{\rm Bo}^2 = 4 g_4 m \nave \; ,
\label{kBo}
\ee
which we will call the Bogoliubov wavenumber.
Here $\nave = \langle \psi^{\dagger} \psi \rangle$ is the average density of 
the gas.
Particles with momenta $p\gg \hbar k_{\rm Bo}$ can be considered as weakly 
interacting, while those with smaller momenta in general cannot.
For $p \ll \hbar k_{\rm Bo}$, elementary excitations of a fully condensed gas
have a linear dispersion law; these are Bogoliubov's phonons \cite{Bogoliubov}.

In addition to the Bogoliubov length $\lambda_{\rm Bo} = 2\pi/ k_{\rm Bo}$, which
is time-independent, let us consider the time-dependent correlation length of 
the gas $\lambda_0(t)$. The correlation length is defined as the characteristic
scale at which the two-point function $\langle \psi^{\dagger} (\x) \psi(0) \rangle$
changes its behavior (e.g. from a power law to an exponential). 
The corresponding wavenumber is $k_0(t) = 2\pi/ \lambda_0(t)$. 
BEC corresponds to 
the growth of the correlation 
length with time, and we now summarize some previous results concerning this
growth \cite{Kraichnan}--\cite{yale}.

If the initial value of the correlation length is small (we say that the initial
sate is disordered),
\be
\lambda_0(0) \ll \lambda_{\rm Bo} \;, 
\label{dis}
\ee
BEC proceeds in two stages. At the first stage, 
$\lambda_0(t) \ll \lambda_{\rm Bo}$.
If one assumes that the main role is played by interactions of particles with
wavenumbers larger than or of order $k_0(t)$, one can use
a quantum Boltzmann equation, because at this stage all such particles are 
weakly interacting.
The solution to the Boltzmann equation \cite{ST} shows that the main feature of 
the spectrum is a turbulent front that propagates to smaller wavenumbers.
The position of the front can be identified with $k_0(t)$. 

When the correlation length
exceeds $\lambda_{\rm Bo}$, the particles begin to interact strongly with
their local environment, and the standard Boltzmann equation cannot be used.
At this second stage the growth of the correlation length
continues through {\em coarsening}, i.e. a process of alignment of the field
on progressively larger spatial scales \cite{KS,yale}, 
characterized by a power-law dependence of $\lambda_0$ on time (at large times):
\be
\lambda_0(t) = {\rm const} \times t^{\alpha} \; .
\label{coar}
\ee
We will return to discussion of the scaling exponent $\alpha$ shortly. Also,
as we will see below, the particle number contained in modes with $k<\kBo$ is
in fact concentrated in the ever-shrinking region $k\sim k_0(t)$. So, at this
stage the correlation length $\lambda_0$ gives the typical size of correlated 
patches (i.e. regions over which the phase of the field is approximately 
constant).

If the initial value of the correlation length is already large (the initial
state is partially ordered),
\be
\lambda_0(0) \gg \lambda_{\rm Bo} \;, 
\label{ord}
\ee
there is no room for a ``preparatory'' stage that can be described by 
a Boltzmann equation for the particles: one has to use a different description
from the beginning. Here we consider only the late-time coarsening
stage, which is present in either of the two cases.

How does one go about determining the scaling exponent $\alpha$?
The fastest the correlation length can possibly grow is at the speed of sound
(cf. ref. \cite{Kraichnan}),
which in our case is the speed of Bogoliubov's phonons
\be
v_s = (g_4 \nave / m)^{1/2} \; .
\label{vs}
\ee
If one replaces (\ref{H}) with a theory of non-interacting phonons, as it is 
done in the ``phase-only'' model of ref. \cite{yale}, the only scaling law one
can possibly discover is $\lambda_0(t) \sim v_s  t$, corresponding 
to $\alpha = 1$. Although this value of $\alpha$ is consistent with the results
of numerical simulations \cite{yale} of the full theory (\ref{H}), the questions 
remain if the approximation of non-interacting phonons is a good one, and why their
free propagation should have anything to do with the growth of the correlation
length. 

In the present paper, we attempt to answer these questions by taking a 
look at what happens in the momentum space, i.e. at the evolution of the power
spectrum of the field. 
We find that a consistent description of the power spectrum is obtained by assuming
that the longest-wave phonons, far from being non-interacting, 
are in the opposite, maximally nonlinear, regime. This assumption leads to
a prediction of an inverse (i.e. directed towards the infrared) turbulent 
cascade of the particle number, with a definite value of the scaling exponent.
This prediction is confirmed by numerical simulations in two and three spatial
dimensions.

We stress that this {\em strong turbulence}, characterized by strong quasiparticle
interactions, is completely different from {\em weak turbulence} found in a 
variety of physical systems on the basis of a Boltzmann equations for weakly 
interacting quasiparticles \cite{ZLF}. Likewise, it is different from the turbulent
solution obtained numerically in ref. \cite{ST} for the Boltzmann stage of the
evolution of a Bose gas.

Because phonons at wavenumbers $k \sim 1/\lambda_0$ are
maximally nonlinear, the rate at which the particle number is transferred to
still larger scales is of order of the phonon frequency $\omega=v_s k$.
The time at which the order is established at a scale $k$ is therefore
$t \sim 1/\omega$, which leads to 
\be
\lambda_0(t) \sim v_s t \; .
\label{lin}
\ee
Thus, the linear scaling law for the correlation length obtains, and the speed
of BEC is of order $v_s$, even though
phonons are by no means free-propagating.

The paper is organized as follows.
In Sect. 2 we present analytical considerations, and in Sect. 3 the numerical
results. Concluding Sect. 4 contains estimates of condensation times 
for trapped atomic gases and galactic dark matter.

\section{Acoustic turbulence}
For a fully ordered Bose gas (i.e. a Bose-Einstein condensate plus small
fluctuations), the spectrum of quasiparticles had been found by Bogoliubov
\cite{Bogoliubov}
for an arbitrary pairwise potential. For the local interaction contained
in (\ref{H}) the dispersion law, $\omega(k)$, is given by
\be
\omega^2(k) = \frac{g_4 \nave}{m} k^2 + 
\frac{{\hbar}^2 k^4}{4m^2} \; .
\label{disp}
\ee
Hence, the wavenumber $k_{\rm Bo}$, eq. (\ref{kBo}), separates those $k$ for which
the spectrum is significantly modified by the interaction with the condensate
from those for which quasiparticles approximately coincide with the ordinary
particles of the gas.

For a partially ordered gas, with the typical size of correlated domains
(the correlation length) equal to some $\lambda_0$, eq. (\ref{disp}) can
be applied to excitations with $k \gg k_0 =2\pi / \lambda_0$, because these 
can be viewed as propagating on an ordered background (we will make the latter
statement more precise shortly). In this section, we will present analytical 
arguments in favor of a certain form of the power spectrum of $\psi$ in the 
low-momentum range
\be
k_0 \ll k \ll k_{\rm Bo} \; .
\label{range}
\ee
Recall that we consider the stage when $k_0 \ll k_{\rm Bo}$.
For wavenumbers in the range (\ref{range}), the quasiparticle dispersion
law $\omega(k)$ is nearly linear in $k$, so we refer to the excitations as
phonons.

In perturbation theory,
one writes $\psi = \sqrt{n} \exp(i\theta/\hbar)$ and
defines the zeroth-order approximation by
replacing $n$ with $\nave$ in the first term in (\ref{H}). This gives
\be
H \approx H_0 = \frac{\nave}{2 m} (\nabla \theta)^2 + \half g_4 n^2 \; .
\label{H0}
\ee 
From the kinetic term in the Lagrangian, 
$i \hbar \psi^{\dagger}\partial_t \psi$, we find that 
$-n$ is the momentum conjugate to $\theta$.
We now see that (\ref{H0}) is a theory of noninteracting waves with the linear
dispersion law
\be
\omega^2(k) = \frac{g_4 \nave}{m} k^2 \; .
\label{disp0}
\ee
This theory coincides with the ``phase-only'' model considered in ref. \cite{yale},
except that we do not introduce any coupling to a heat bath. Eq. (\ref{disp0})
is, of course, the low-momentum limit of (\ref{disp}).

Next, allow $n$ to deviate from $\nave$ in the gradient term and consider
small $\delta n = n - \nave$. To the second order in $\delta n$, the Hamiltonian
becomes $H \approx H_0 + H_1$, where
\be
H_1 = \frac{\delta n}{2 m} (\nabla \theta)^2 
+ \frac{\hbar^2}{8m} \frac{(\nabla n)^2}{\nave} \; .
\label{H1}
\ee
The second term results in a correction to the dispersion law, restoring the
full eq. (\ref{disp}). (A similar way of deriving Bogoliubov's spectrum 
was used in ref. \cite{KSS}.)
The first term is an interaction among quasiparticles.
Using the  decomposition of $\theta$ and $\delta n$ in terms of quasiparticle
creation and annihilation operators,
\ba
\theta & = & {1\over \sqrt{V}} \sum_{\k} \sqrt{\frac{\hbar g_4}{2\omega_k}}
\left( a_{\k} + a^{\dagger}_{-\k} \right) \e^{i\k\x} \; , \\
\delta n & = & {i\over \sqrt{V}} \sum_{\k} \sqrt{\frac{\hbar\omega_k}{2g_4}}
\left( a_{\k} - a^{\dagger}_{-\k} \right) \e^{i\k\x} \; ,
\ea
where $V$ is the total volume, we find that the matrix element of the 
interaction, $U(\k_1, \k_2, \k_3)$, has the conventional phonon 
form at low momenta:
\be 
U(\k_1, \k_2, \k_3) \propto \left( k_1 k_2 k_3 \right)^{1/2} \; .
\label{int}
\ee

We can now understand better the sense in which we can talk about 
quasiparticles in a partially
ordered state. First, note that in addition to perturbative excitations there are
nonperturbative ones---vortices in two dimensions and vortex lines in three.
These cannot be described in terms of quasiparticles because $n$ vanishes at
the core of a defect, making it impossible to even define $\theta$ there.
We expect that the typical distance between the defects is of order of the
correlation length $\lambda_0$ (cf. ref. \cite{KS}). 
Because perturbative excitations scatter strongly 
on the defects, the quasiparticle concept breaks down for $k \leq k_0$: such 
excitations do not have any room to propagate. Of course, excitations with
$k \gg k_0$ also scatter on the defects, but only on scales of order $\lambda_0$.
On shorter scales, they can be treated as if the defects were absent.
In particular, if the interaction among quasiparticles themselves leads to a
mean-free path shorter than $\lambda_0$, it is that interaction (rather than 
scattering off the defects) that determines the quasiparticle lifetime.
We will now argue that such is indeed the case.

Indeed, it is precisely for acoustic (close to linear)
dispersion laws that the
conditions under which quasiparticle interactions can be treated perturbatively
become quite subtle.
The reason has been explained in ref. \cite{explain}
and is the following. For an exactly linear
dispersion law, wave packets do not disperse, so if two excitations are produced
with nearly parallel momenta, they will stay together for a long time, 
breaking
the usual condition of applicability of perturbation theory.
In our case, the dispersion law $\omega(k)$, given by eq. (\ref{disp}), is not
exactly linear but curves upward from a straight line. As a result, the
process in which two quasiparticles merge into one is kinematically allowed
but, because the deviation of the dispersion from linear is small at small $k$,
the initial momenta have to be almost parallel. Thus, quasiparticles are forced
into the kinematic arrangement in which the interaction between them can be large.
In particular, a {\em weak turbulence} spectrum, obtained as a solution to a
quantum Boltzmann equation for phonons 
(with an interaction of the form (\ref{int})),
is applicable only at sufficiently large $k$ \cite{applic}.
Therefore, in the small $k$ region (\ref{range}), we will be searching for
{\em strong turbulence}, for which the lifetime of a quasiparticle is of order
of its inverse frequency, and the Boltzmann equation cannot be applied.

\begin{figure}
\leavevmode\epsfysize=3.0in \epsfbox{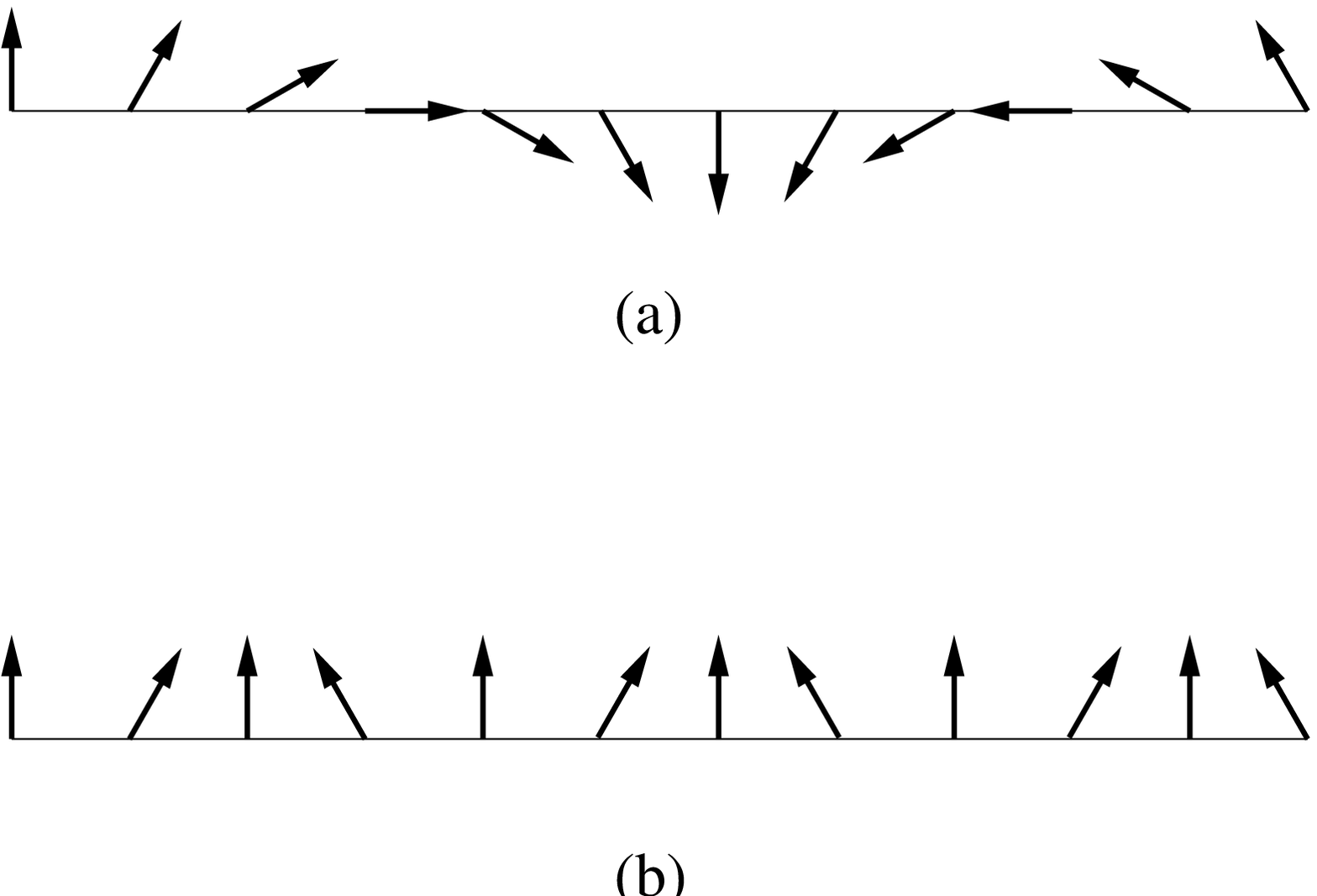}
\vspace*{0.2in}
\caption{Difference between uncondensed (a) and condensed (b) states with
equal energies and equal particle numbers.
}
\label{fig:arrows}
\end{figure}

To see further the importance of the scattering process in which two quasiparticles
merge into one, let us visualize BEC as going from the situation shown in 
Fig. \ref{fig:arrows}a to that shown in Fig. \ref{fig:arrows}b. 
The field $\psi$ is represented by arrows
on a grid, with the length of an arrow corresponding to $|\psi|$, and the angle
of rotation to $\arg\psi \propto \theta$. The grid is made one-dimensional for
illustrative purposes only; in fact, we do not consider $d=1$ in this paper.
We have chosen these particular configurations in such a way that
the lengths of all arrows are about the same, and the magnitudes of angular
differences between neighboring arrows are also the same, for all sites
in Figs. \ref{fig:arrows}a and \ref{fig:arrows}b. Thus, both the gradient and 
potential energies are about the
same for the two figures, i.e. the energy is conserved. Nevertheless, the state
of Fig. \ref{fig:arrows}a has no long-range order, while the state of 
Fig. \ref{fig:arrows}b clearly has some.
We can also say that the first state has many low-momentum
quasiparticles, while the second state has relatively few high-momentum ones. The
transition from the first state to the second should therefore occur through
merging of quasiparticles, leading to a decrease in their number and an increase in
the average quasiparticle energy.

We now see that the process of BEC is largely analogous to turbulence in an
incompressible fluid. Instead of turbulent eddies we now have quasiparticles,
and instead of a larger eddy splitting into smaller ones we now have two quasiparticles
merging into one (of course the inverse process exists as well but, as we have
seen, for BEC to occur the merging of quasiparticles should statistically 
prevail over the splitting).
Moreover, we have seen that the interaction among quasiparticles in this process
has every chance to be strong, so that the corresponding mean-free path can be 
short---shorter than the correlation length $\lambda_0$. If that is indeed so,
this process is the main interaction effect for the entire momentum range 
(\ref{range}), so this range becomes analogous to the inertial range in an 
incompressible fluid (cf. ref. \cite{Kraichnan}). We can therefore try to deduce 
the power spectrum of $\psi$
by an argument a la Kolmogorov and Obukhov (as reviewed for instance in the
Landau-Lifshitz textbook \cite{LL}).

Unlike the case of an incompressible fluid, where only the energy is conserved,
in our case there are two conserved quantities: the energy and
the number of particles (by particles we mean here the ordinary particles of 
the gas, as opposed to quasiparticles, whose number is not conserved). The process
of BEC can be viewed either as particle number accumulating in low-momentum modes
or, as we have seen above, as energy accumulating in high-momentum modes. So, we
have to decide with respect to which quantity the turbulent cascade will be
established. A posteriori checks show that the only consistent choice
is an inverse (i.e. directed towards the infrared) cascade of the particle number.

We therefore proceed to define the particle number contained in all scales 
shorter than a certain scale $\lambda = 2\pi / k$, with $k$ in the inertial range 
(\ref{range}):
\be
N(k,t) = \sum_{|\k'|>k} \langle \psi^{\dagger}_{\k'} \psi_{\k'} (t) \rangle \; ,
\label{Nk}
\ee
where the angular brackets denote averaging over the quantum state of the system.
A turbulent cascade corresponds to a constant ($k$-independent) 
flux of the particle number in
momentum space. The flux can be estimated by multiplying $N(k)$ by the rate at
which quasiparticle interactions transport the particle number to larger scales. 
This rate is simply the inverse 
quasiparticle lifetime; for strong turbulence it is of order of the quasiparticle
frequency $\omega(k) \approx v_s k$. Thus, we obtain
\be
k N(k,t) = {\tilde A}(t) \; ,
\label{A}
\ee
where ${\tilde A}$ is a $k$-independent function of time. Eq. (\ref{A})
corresponds to the following power spectrum of $\psi$
\be
P(k) = \langle |\psi_{\k}|^2 \rangle = \frac{A(t)}{k^{d+1}} \; ,
\label{pws}
\ee
where $A$ is proportional to ${\tilde A}$, and $d=2$ or 3 is the number of spatial 
dimensions. 

We can now consider some consistency checks. According to the previous discussion,
we expect that the particle number, given by the integral over $\k$ of $P(k)$
saturates in the infrared, while the gradient energy, given by the integral of
$k^2 P(k)$ saturates in the ultraviolet. Both conditions
indeed hold for the power spectrum (\ref{pws}). 

The integrals in question are cut-off at the boundaries of the range
(\ref{range}). In particular, the integral corresponding to the total
particle number is cut-off at $k \sim k_0$:
\be
N_{\rm tot} \sim \frac{A(t)}{k_0(t)}  V \; ,
\label{Ntot}
\ee
where $V$ is the total volume.
The time dependence of $k_0$ can be found by pushing our estimate for the quasiparticle
lifetime $\tau(k) \sim 1/v_s k$ to the limit of its applicability, i.e. to $k\sim k_0$.
Then, $\tau(k)$ can be interpreted as the timescale at which the turbulent spectrum
reaches the wavenumber $k$, or equivalently the size of correlated domains extends to
spatial scale $\lambda = 2\pi / k$. The estimate $\tau(k_0) \sim 1/v_s k_0$
corresponds to 
\be
\lambda_0(t) \sim v_s t \; ,
\label{lam0}
\ee
i.e. a linear growth of the correlation length with time.
We stress that this growth
is due to a strong interaction among quasiparticles and cannot be described by 
the phase-only model (\ref{H0}), which contains no such interaction at all.

For comparison, in the ring-model approximation for $d=3$
Kraichnan \cite{Kraichnan} finds $P(k) \propto k^{-13/3}$ (which
is curiously close to our scaling (\ref{pws})), and
$\lambda_0(t) \propto \sqrt{t}$. (Note that notations differ: our
$k_0$ corresponds to Kraichnan's $k_*$, and our $\hbar \kBo^2/ 2m$ is twice 
his $\omega_0$.)

Because the total number of particles is time-independent, the scaling
$k_0 \propto 1/t$ implies, through (\ref{Ntot}), that
\be
A(t) = \frac{{\rm const}}{t} \; .
\label{At}
\ee
Thus, the spectrum propagates to smaller $k$ according to (\ref{lam0}), while
the magnitude of the spectrum at a given $k$ decreases according to (\ref{At}).

We wish to stress again that the power spectrum (\ref{pws}) describes the distribution
of ordinary particles of the gas, as opposed to quasiparticles. This distinction
should be kept in mind when making comparisons with the literature.
For instance, weak turbulence is often described by a distribution of quasiparticles
\cite{ZLF}, 
the same ones for which the Boltzmann equation is written. In our case, a transition 
to quasiparticle spectra may be difficult: due to the strong interactions,
the statistics of quasiparticles is likely to be non-Gaussian and therefore not
amenable to a simple description in terms of occupation numbers.

\section{Numerical results}
The results of the previous section are all based on one central assumption, namely,
that a turbulent cascade, described by a constant flux of particles in the $k$-space,
is indeed established. As well known in other contexts, a turbulent cascade can only 
occur if the interaction (in our case, among quasiparticles) is in a certain 
sense local in the $k$-space. In our case, that means that the main role should be 
played by interactions among quasiparticles whose momenta are of the same order of 
magnitude, rather than, say, by processes in which a quasiparticle will jump directly 
to a momentum of order $\kBo$.

Another ingredient of our theory is the use of the maximally nonlinear estimate
for the quasiparticle lifetime.  Although we tried to motivate this estimate in 
the previous section, it should still be properly considered as an assumption.

For cases of weak turbulence, the assumption of locality can be verified, and
the quasiparticle lifetime can be found
by using the solution to the corresponding Boltzmann equation \cite{ZLF}. 
We are not aware of
any such analytical method for strong turbulence. So, we have resorted to
numerical simulations. The test of the assumptions will be indirect---through the
form of the power spectrum.

Operationally, the numerical approach is quite straightforward.
One assumes that the pile-up of particles at low momenta leads to large
occupation numbers in low-momentum modes, so that the field $\psi$ can be regarded
as classical and described by a classical equation of motion (which for
the Hamiltonian (\ref{H}) is known as the Gross-Pitaevsky equation).
The classical equation can be applied to
the coarsening stage of the evolution \cite{KS} (as well as to the Boltzmann one
\cite{Svistunov}), as long as the assumption of classicality is good.
This approach was used in the previous numerical study of BEC
\cite{yale}, but the form of the power spectrum was not addressed there. 
(For attractive interactions, the classical approximation was used in refs. 
\cite{dew,grinst}.) Here we concentrate on the power spectrum of $\psi$.

The condition for the occupation numbers to be large, though, 
is not entirely trivial. For wavenumbers in the range (\ref{range}), 
it can be derived if we use the already obtained power spectrum 
(\ref{pws}) (anticipating that this form of the spectrum will be confirmed
numerically). Defining the particle occupation numbers
as $n_\k = \langle \psi^{\dagger}_\k \psi_\k\rangle$ and using (\ref{Ntot}), 
we find
\be
n_\k \sim \frac{\nave k_0(t)}{k^{d+1}}  \; .
\label{nk}
\ee
Thus, at any given $k$ the occupation number actually {\em decreases} with time.
Still, for all $k$ from the range (\ref{range})
\be
n_\k \gg \frac{\nave k_0(t)}{\kBo^{d+1}} 
= \frac{k_0(t)}{\epsilon k_{\rm Bo}} \; ,
\label{eps}
\ee
where $\epsilon = \kBo^d / \nave$.
So, if the interaction parameter 
$\epsilon$ is small, the condition $n_\k \gg 1$ can hold even for an
already small ratio $k_0 /\kBo$.

Furthermore, the use of the classical approximation implies that $n_{\k} \gg 1$
for all  ``important'' momenta. That is certainly not so for a thermal
state, in which both the particle number and the energy are concentrated in modes
with $n_{\k} \sim 1$. For this reason, we
will consider nonthermal initial conditions. In addition, we will have to make
sure that the high-momentum modes do not get overly populated in the course of
the evolution. This latter requirement is rather nontrivial, since the gradient
energy contained in the range (\ref{range}) is, for the power spectrum (\ref{pws}),
of order
\be
\int_{k_0}^{\kBo} d\k k^2 P(k) \sim A(t) \kBo \; ,
\label{gren}
\ee
and thus decreases with time.
So, unlike the particle number, the gradient energy is not contained in 
the low momentum modes
but is being transferred to modes with $k > \kBo$. In this way, the low-momentum
part of the gas gets ``cooled down'', while the high-momentum part is ``heated up''.
In a quantum system, the flow of energy to high momenta leads to thermalization,
but in a classical system on a lattice, the energy eventually becomes dominated
by the cutoff modes. When that happens, the simulation breaks down.
Fortunately, when we plot the power 
spectra, we will be able to monitor their high-momentum content as well. 

After these remarks on the applicability of the classical approximation, we
proceed to solve
the classical equation of motion for a dilute Bose gas (the
Gross-Pitaevsky equation), which after some rescalings takes the form
\be
i\frac{\partial \psi}{\partial t} = -\nabla^2 \psi + |\psi|^2 \psi \; .
\label{eqm}
\ee
The rescalings that lead from (\ref{H}) to (\ref{eqm}) are as follows. A factor
of $\sqrt{g_4}$ is absorbed into $\psi$, a factor of $\hbar$ into the time derivative,
and a factor of $\hbar / \sqrt{2m}$ into the spatial derivatives. We hope that using
the same letters for the rescaled quantities as for the original ones
will cause no confusion, as only the rescaled quantities are used throughout this 
section. The rescalings correspond formally to setting $\hbar=1$, $g_4 = 1$, and
$m=1/2$. In particular, the expression for the Bogoliubov wavenumber becomes
\be
\kBo = \sqrt{2\nave} \; .
\label{kBor}
\ee
The power spectrum of a classical field is defined by averaging over
direction of the wavevector:
\be
P(k) = \frac{1}{\Omega_{\rm tot}} \int d\Omega_{\k} |\psi_{\k}|^2 \; ,
\label{pwscl}
\ee
where $\Omega_{\rm tot} = 2\pi$ in $d=2$, and  $\Omega_{\rm tot} = 4\pi$ in $d=3$.
It satisfies the equality
\be
\nave = (2\pi)^{-d} \int d^d k  P(k) \; .
\ee

Eq. (\ref{eqm}) was solved in two- and three-dimensional boxes will equal sides
of length $L$ and periodic boundary conditions. 
We used an operator-splitting
algorithm, based on alternation of two different types of updates corresponding to
the individual terms on the right-hand side of (\ref{eqm}). The updates were
alternated so as to maintain the second-order accuracy in time. The action
of $\nabla^2$ was computed in the momentum space, using the fast Fourier transform.
The algorithm conserves the number of particles exactly. Energy nonconservation,
for the times shown in the figures, was about 0.2\% in $d=3$ and about 2\% in
$d=2$, relative to the value of energy at $t=0$.

The initial condition was taken in the form
\be
\psi_\k(t=0) = \sqrt{P(k,0)} \exp(i\gamma_\k) \; ,
\label{init}
\ee
with random uniformly distributed phases $\gamma_\k$ and a nonthermal power spectrum
\be
P(k,0) = B \exp(-k^2 / \mu^2) \; .
\label{Pini}
\ee
For both runs shown in the figures, we used $\mu = \pi$, while the values of $B$
were such that the average densities equaled
$\nave = 31.4$ for $d=2$, and $\nave =27.8$ for $d=3$, corresponding to $\kBo = 7.93$
and $\kBo = 7.46$. The parameter $\mu$ can be viewed as the wavenumber corresponding
to the initial correlation length $\lambda_0(0)$: $\mu = \pi$ means
$\lambda_0(0) = 2$. Because $\mu < \kBo$ for both runs, the gas was in 
the coarsening regime from the beginning.

In $d=2$, a Bose gas cannot support long-range order at any finite temperature
\cite{Hohenberg}.
However, a classical simulation of an isolated
system described by eq. (\ref{eqm}), with a finite total energy and a large 
(ideally, infinite)
ultraviolet cutoff, corresponds to an effectively zero temperature. In particular,
at a finite temperature there would be a finite equilibrium density of vortices
(or vortex-antivortex pairs). In a classical simulation, the number of vortices and 
antivortices decreases with time.

\begin{figure}
\leavevmode\epsfysize=3.0in \epsfbox{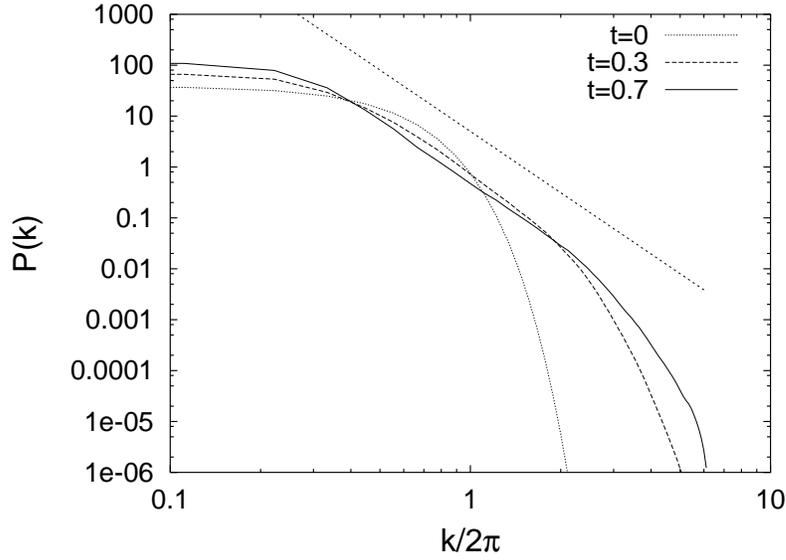}
\vspace*{0.2in}
\caption{Power spectrum of $\psi$ in $d=3$ at three different moments of time,
obtained on a $64^3$ lattice with side length $L=9$. The straight line is a
$k^{-4}$ power law.
}
\label{fig:3d}
\end{figure}

\begin{figure}
\leavevmode\epsfysize=3.0in \epsfbox{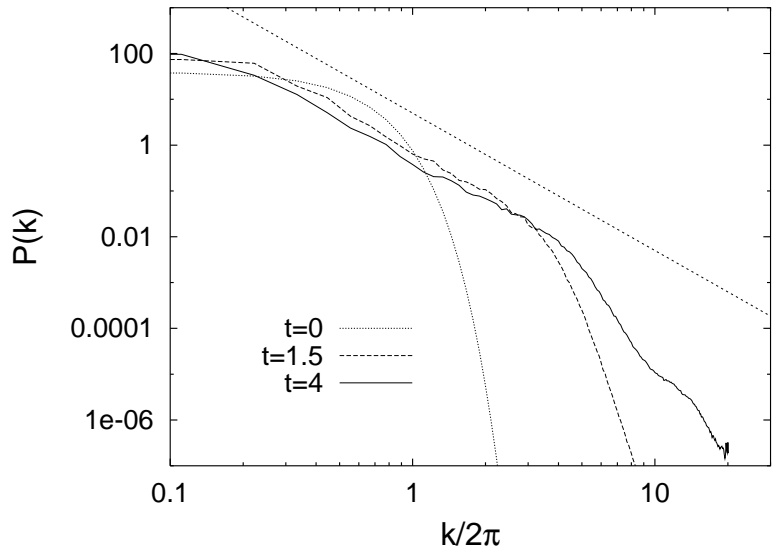}
\vspace*{0.2in}
\caption{Power spectrum of $\psi$ in $d=2$ at three different moments of time, 
obtained on a $256^2$ lattice with side length $L=9$. The straight line is a
$k^{-3}$ power law.
}
\label{fig:2d}
\end{figure}

We show the evolution of the power spectrum for $d=3$ in Fig. \ref{fig:3d}, 
and for $d=2$ in Fig. \ref{fig:2d}. 
We see that in either case a region of 
power law forms, well approximated by eq. (\ref{pws}). 
According to our previous
discussion, the power law should extend in the ultraviolet to wavenumbers of
order $\kBo$. Numerically, we find a ``knee'' in the power spectrum at
$k\approx 2\kBo$.

It is interesting to compare our results to those obtained in ref. \cite{DF}
for an externally driven $d=2$ Gross-Pitaevsky equation. In that case, 
the total number of particles is not conserved, so the parameters $\kBo$ 
and $v_s$ are time-dependent. Nevertheless, if the change of these parameters 
is sufficiently slow, we can still expect formation of an inverse cascade 
via quasiparticle scattering.\footnote{
It is not immediately clear what the rate of 
change of $N_{\rm tot}$ should be compared to. However, for the case called 
moderate pumping in ref. \cite{DF}, this rate is much smaller than even
the smallest energy 
scale present in the Hamiltonian (\ref{H}), $\hbar^2 k_{\min}^2 / 2m$, 
where $k_{\min}$ is 
the minimal nonzero lattice wavenumber. It seems plausible that in this case
the quasiparticle spectrum and the scattering rates are not affected too much by 
the time-dependence of $N_{\rm tot}$.
} 
The authors of \cite{DF} obtain a power-law spectrum
$n_{\k} \propto k^{-2}$, i.e. an index different from ours, but they also 
observe that the spectrum (including the index) is anisotropic. We believe 
the anisotropy can be
explained by the presence of some remaining vortex-antivortex pairs, perhaps
just one such pair in the entire integration volume. Such well-separated vortices
are easily pinned on a lattice, so a single pair can survive 
for a long time, causing an effective anisotropy. Then, it makes sense to average
the power spectrum over the direction of the 
wavenumber, as we do in our definition (\ref{pwscl}). In the presence of anisotropy,
the main contribution to the average will come from directions in which 
fluctuations are the largest. It is precisely in such directions that the spectrum
obtained in ref. \cite{DF} is very similar to ours, with a relatively flat region at
the smallest $k$, and a power law $n_{\k} \propto k^{-3}$ at larger $k$.

The spatial sizes of our lattices were not sufficient to perform quantitative
checks of the temporal scaling laws (\ref{lam0}) and (\ref{At}). While
for the first of these one may be content with the results of ref. \cite{yale},
it would certainly be interesting to directly test (\ref{At}) on larger lattices.
Qualitatively, the predicted by (\ref{At})
decrease in the magnitude of the power spectrum in the inertial range
is clearly seen in both Fig. \ref{fig:3d} and Fig. \ref{fig:2d}.

\section{Discussion}
We now apply our results to two cases: (i) BEC in atomic vapors and (ii) 
formation of a Bose-Einstein condensate from galactic dark matter.
In both cases, we estimate the speed of BEC as being of order of the sound
velocity (\ref{vs}). We can then estimate the time scale required for BEC
to complete on physically interesting spatial scales.

Using the estimate $4\times 10^{14}$ cm$^{-3}$, obtained in ref. \cite{exp3}
for the number density of sodium atoms, as $\nave$ in eq. (\ref{vs}) and noting 
that our
parameter $g_4$ is identical to ${\tilde U}$ used in that paper, we find 
$g_4 \nave = 520 ~{\rm nK}$, and
\be
v_s = 1.4 ~{\rm cm/s} \; .
\label{vsgas}
\ee
Thus, for a gas cloud a few micron across, we estimate that
the coarsening time is shorter than 1 ms. The duration of
evaporative cooling in ref. \cite{exp3} was 7 s, so the coarsening time was
not a limiting factor in that experiment. In the experiment with 
rubidium \cite{exp1}, density was about two orders of magnitude smaller, which
reduces $v_s$ by about an order of magnitude compared to (\ref{vsgas}) ($g_4$
and $m$ are also different, of course, but by smaller factors). Still,
the coarsening time remains much smaller than the time scale of evaporative 
cooling.

On the other hand, the relaxation time corresponding to the first, Boltzmann,
stage of BEC in these gases is quite large. For a gas of lithium-7 atoms 
(which have an attractive interaction at low densities), with the parameters of
the experiment of ref. \cite{exp2}, this time was estimated in ref. \cite{dew}
to be of order 1 s. Similar estimates can be obtained for gases with repulsive
interactions. First, form the dimensionless parameter of nonlinearity
\be
\xi_T = \frac{4\pi\hbar^2 a n}{m T} \; ,
\label{xiT}
\ee
where $a$ is the scattering length, and $T$ is the temperature in energy units. 
For rubidium-87,
the authors of ref. \cite{exp1} quote $a \sim 10^{-6}$ cm, $n=2.5\times 10^{12}$
cm$^{-3}$, and $T = 170$ nK. With these values of the parameters, 
$\xi_T \sim 10^{-2}$. In the Boltzmann approximation, the relaxation rate is 
proportional to $a^2$. Hence, we can roughly estimate the relaxation time as
\be
t_r \sim \frac{\hbar}{T \xi_T^2} \sim 0.4~{\rm s} \; .
\label{tr}
\ee
The authors of \cite{exp1} noted that after reaching the condensation 
temperature the condensate fraction grew for 2 seconds. Given how crude the
estimate (\ref{tr}) is, it seems plausible that this delay can be attributed
to the Boltzmann relaxation.

The distinction between disordered and partially
ordered initial states is important also for a Bose gas with an attractive 
two-particle interaction (a negative $a$). The wavenumber
$\kBo$ is now defined (in units with $\hbar =1$) by
\be
 k_{\rm Bo}^2 = 4 |g_4| m \nave = 16 \pi |a| \nave 
\label{kBo_attr}
\ee
(the last equality is specific for three spatial dimensions). 
The quadratic dependence of the rate of collapse (fragmentation
into individual drops) on the scattering length \cite{dew} applies 
only when the initial state is disordered, 
i.e. the correlation length $\lambda_0$ is smaller than $\lambda_{\rm Bo}$.
In the opposite limit of a large correlation length, collapse 
in a sufficiently dense system will occur by linear instability. 
Indeed, the dispersion law for excitations with $k \gg k_0$ is now
\be
\omega^2(k) = \frac{1}{4m^2} \left( k^4 - \kBo^2 k^2 \right) \; ,
\label{disp_attr}
\ee
which corresponds to an instability for all $k < \kBo$.
The instability is the strongest for $k = \kBo/ \sqrt{2}$, where its rate
is
\be
|\omega|_{\max} = \kBo^2 / 4m \; .
\label{ommax}
\ee
A large initial correlation length can be achieved, for example, 
if the scattering
length was originally positive, but after some time was switched to a negative
value, as in a recent experiment \cite{collapse}. If we imagine that the gas
is confined in a sphere of radius $R$ with Dirichlet boundary conditions for
$\psi$, and we gradually increase $\kBo$, by increasing $|a|$ or $\nave$, 
the instability will first appear when $\kBo = \pi / R$. Then, after $\kBo$
passes a transitional region with $\kBo \sim \pi / R$, the time scale of collapse
will, as seen from (\ref{ommax}), become inversely proportional to the 
scattering length and to the density of atoms. In our estimation, 
these scaling laws---especially the former---are well born out in the experimental 
results of ref. \cite{collapse}.

For Bose gases considered in cosmological and 
astrophysical applications, one should take into account gravitational effects. 
If the particles are quanta of a real scalar field
with a potential
\be
V(\phi)= \half m^2 \phi^2 + {\lphi \over 4} \phi^4 \; ,
\ee
then, in the absence of gravity, (\ref{H}) is the correct nonrelativistic 
limit, provided
\be
g_4 = \frac{3\lphi}{4 m^2} \; .
\ee
This limit is obtained by writing 
\be
\phi = \frac{1}{\sqrt{2m}} \left( \psi \e^{-imt} + \psi^{\dagger} \e^{imt} \right)
\ee
and assuming that $\psi$ changes with a frequency scale much smaller than $m$.
Eq. (\ref{vs}) for the sound velocity becomes
\be
v_s = \left( \frac{3\lphi \rho}{4m^4} \right)^{1/2} \; ,
\ee
where $\rho = m\nave$ is the mass density. 

Gravitational interaction (in Newtonian gravity)
modifies the quasiparticle dispersion law, so that now
\be
\omega^2(k) = \frac{1}{4m^2} \left( k^4 + \kBo^2 k^2 - k_J^4 \right) \; ,
\label{disp_grav}
\ee
where $k_J^4 = 16 \pi G m^2 \nave$ is the Jeans scale of the gas. (For a
discussion of various length scales in a self-gravitating Bose gas, and 
various types of gravitational instabilities, see ref. \cite{grinst}.)
In a partially ordered Bose gas, eq. (\ref{disp_grav}), similarly to 
eq. (\ref{disp}), applies only at scales smaller than the correlation length 
$\lambda_0(t)$. It follows that
once $\lambda_0$ exceeds a critical value $\lmax$,
\be
\lmax^2 = \frac{4\pi^2 \kBo^2}{k_J^4} = \frac{3\pi \lphi}{4G m^4} \; ,
\label{lmax}
\ee
the gas becomes gravitationally unstable via a linear (Jeans) instability 
and fragments into smaller clumps. Similar expressions were obtained 
in \cite{it86}, \cite{Peebles}--\cite{Riotto&Tkachev} for the radius 
of a stable gravitationally bound object.

According to the general theory presented above, BEC at scale $\lmax$ will
occur through coarsening only if $\lmax$ is larger than the Bogoliubov
wavelength $\lambda_{\rm Bo}$, or equivalently $\kBo > k_J$. This corresponds
to
\be
\lphi > \frac{4 m^3}{3} \left( \frac{\pi G}{\rho} \right)^{1/2} = 
10^{-25} \left(\frac{m}{1~{\rm eV}} \right)^3 \; .
\label{bound_coars}
\ee
In the last estimate, we have used
$\rho = 0.02 M_{\odot} / {\rm pc}^3$ \cite{Firmani&al} for the mass density 
at a galaxy core. The time scale $\tau$ required for 
coarsening to extend to scale $\lmax$ can be estimated as
\be
\tau = \frac{\lmax}{v_s} = \left(\frac{\pi}{G\rho}\right)^{1/2} 
=  2 \times 10^8 ~{\rm yrs} \; .
\label{tau_coars}
\ee
Thus, regardless of the coupling, 
$\tau$ is much smaller than the current age of the universe (although much larger 
that the age at the time of galaxy formation).

On the other hand, the condition for the first, Boltzmann, stage of BEC to complete, 
obtained in \cite{it91,Riotto&Tkachev} is, for the same value of $\rho$,
\be
\lphi > 10^{-15} \left(\frac{m}{1~{\rm eV}} \right)^{7/2} \; .
\label{bolt}
\ee
If this condition is satisfied, completion of BEC on scale $\lmax$ is guaranteed by
(\ref{tau_coars}). So, for the gas to be at least linearly stable on the scale 
of the galaxy core, $\lmax$ should exceed the core diameter $2 r_c$. In view of
(\ref{lmax}), this requires
\be
\lphi > 3 \times 10^{-4} 
\left(\frac{r_c}{1~{\rm kpc}}\right)^2 \left(\frac{m}{1~{\rm eV}} \right)^4 \; .
\label{lam}
\ee
Except for very small masses, (\ref{lam}) is a stronger condition
than (\ref{bolt}) for all scales of relevance to galactic dark matter, 
$r_c > 1~{\rm kpc}$. If $\lphi$ satisfies (\ref{bolt}) but not
(\ref{lam}), then the coarsening eventually sets off a Jeans instability, and
the core fragments into smaller objects (Bose stars).

In conclusion, we have presented analytical arguments in favor of strong turbulence
of the particle number occurring at the final, coarsening stage of Bose-Einstein
condensation. We have found the form of the spectrum in the region of
wavenumbers corresponding to an inverse cascade of the particle number. We have
confirmed both the existence and the scaling exponent of the power-law spectrum
by numerical simulations of classical fields in two and three dimensions. According
to our theory, quasiparticles (Bogoliubov's phonons) are in the
maximally nonlinear regime, implying that the correlation length 
grows linearly in time, with the speed of order of the speed of phonons. 
Finally, we have applied these findings to estimate
the time scale of BEC in atomic vapors and in galactic dark matter.

I would like to thank I. Tkachev for getting me interested in theory of BEC and
many useful discussions. Part of this work was done
during the 1999 miniprogram ``Nonequilibrium Quantum Fields'' at the
Institute for Theoretical Physics, Santa Barbara. I thank ITP for hospitality. 
This work was
supported in part by the U.S. Department of Energy through Grant DE-FG02-91ER40681 
(Task B).

\end{document}